\documentstyle[prl,aps,preprint]{revtex}
\begin{document}
\draft

\tighten

\newcommand{\Rn}{{\rm I\!R}} 
\newcommand{\Nn}{{\rm I\!N}} 
\newcommand{\Cn}{{\setbox0=\hbox{
$\displaystyle\rm C$}\hbox{\hbox
to0pt{\kern0.6\wd0\vrule height0.9\ht0\hss}\box0}}} 

\newcommand{\Tr}{{{\bf Tr}}}
\newcommand{\Flim}{{{\lim _{\F_0}\, }}}
\newcommand{\half}{{{1\over 2}}}

\newcommand{\ot}{{\otimes}}
\newcommand{\Hl}{{\cal H}}
\newcommand{\A}{{\cal A}}
\newcommand{\qA}{{\bf A}}
\newcommand{\E}{{\cal E}}
\newcommand{\com}{{C(\Omega)}}
\newcommand{\pa}{{\partial}}
\newcommand{\Bi}{{{\cal B}({\cal H}_i )}}
\newcommand{\B}{{{\cal B}({\cal H})}}
\newcommand{\cB}{{{\cal B}}}
\newcommand{\K}{{\cal K}}
\newcommand{\cL}{{\cal L}}
\newcommand{\pn}{{{1 \over {\sqrt n}}}}
\newcommand{\p}{{{\pi_{\varphi}}}}
\newcommand{\s}{{{\varrho^{{1\over 2}}}}}
\newcommand{\si}{{{\varrho^{{1\over 2}}_i}}}
\newcommand{\sig}{{{\sigma^{{1\over 2}}}}}
\newcommand{\St}{{\cal S}}
\newcommand{\cW}{{\cal W}}
\newcommand{\vp}{{\varphi}}
\newcommand{\M}{{\cal M}}
\newcommand{\om}{{\Omega}}
\newcommand{\cP}{{\cal P}}
\newcommand{\1}{{\bf 1}}


\title{ Does quantum chaos exist? \\
A quantum Lyapunov exponents approach. }

\author{  Adam W. Majewski}

\address{ Institute of Theoretical Physics and Astrophysics\\
University of Gda\'nsk, 80--952 Gda\'nsk, Poland\\ 
e-mail: fizwam@univ.gda.pl}

\maketitle

\begin{abstract}
We shortly review the progress in the domain of deterministic
chaos for quantum dynamical systems.
With the appropriately extended definition
of quantum Lyapunov exponent we analyze various quantum dynamical maps.
It is argued that, within Quantum Mechanics, irregular evolution
for properly chosen observables can coexist with regular and 
predictable evolution of states.

\end{abstract}


\newtheorem{defn}{Definition}
\newtheorem{defns}[defn]{Definitions}
\newtheorem{rem}{Remark}
\newtheorem{rems}[rem]{Remarks}
\newtheorem{lem}{Lemma}
\newtheorem{cor}{Corollary}
\newtheorem{prop}{Proposition}
\newtheorem{theo}{Theorem}
\newtheorem{examp}{Example}

\section{Introduction.}
The theory of deterministic chaos has a venerable
history, dating back to Poincare celebrated work (see \cite{Poincare}).
Here, deterministic chaos refers to deterministic
development with chaotic outcome.
Another way to say this is that from moment to moment
the system is evolving in a deterministic way, i.e., the 
current state of a system depends on the previous state in a rigidly 
determined way. However, measurements made on the system
do not allow the prediction of the state of the system
even moderately far into the future.
It was observed that whenever dynamical chaos
was found, it was accompanied by nonlinearity (cf. \cite{Gutz}).
Another observation made long
time ago was that exponential sensibility in 
nonlinear {\it classical} systems can lead to exceedingly
complicated dynamical behavior
(cf. \cite{Lorenz}). The characteristic features of 
such a behavior, e.g. deterministic unpredictability, positive 
Lyapunov exponents, are typical attributes of chaotic (deterministic) 
dynamical systems (cf. \cite{Gutz}, \cite{ER}). 
We emphasize that the theory of such classical dynamical
systems is a well developed subject and contains
many (concrete) examples of chaotic systems.
For instance, we can describe the dynamics of a classical
system as the evolution in the phase space governed by the
canonical equations of motion.
Solutions of these equations define the
family of trajectories on the phase space.
The complex behavior of certain class
of trajectories can be reflected by the positiveness of
Lyapunov exponents
and for some choices of the Hamiltonians one can
find concrete examples of such chaotic systems (see \cite{Gutz}).

However, contemporary science has been founded 
on {\it quantum mechanics}. Therefore, it is quite natural to
look for the quantum-mechanical description of (classically) chaotic
systems (see e.g. \cite{Alicki}, \cite{Benatti}, \cite{Casati}, 
\cite{Haake}, \cite{Gutz}, \cite{Majew}, \cite{SlomZycz},
\cite{Thirring}, \cite{Toda}, \cite{V1}, \cite{Zyczkowski}). 
To achieve this objective it is important to realize that
the chaotic behavior arises from nonlinearities in the equations of 
motion. Let us remind that the Schr\"odinger equation of motion is 
a linear one. Furthermore, the Koopman's construction, in the 
theory of abstract dynamical systems, leads to linear 
evolution operators (see e.g. \cite{AA}). On that basis, the long 
debate on the existence of quantum chaos leads to a conjecture that 
quantum mechanics suppresses chaotic behavior in such a way
that there is only a room for ``fingerprints" of chaos \cite{B}.

Our objective in this paper is
to show that {\it quantum mechanics
admits, in some special cases, certain features of chaotic behavior.}
To understand the origin of this phenomenon 
let us remind that the dynamics of a classical
system can be described either as the evolution
in the phase-space by the conventional
Hamilton's equations or by the Liouville equations.
Now if we consider a system with some ergodic
properties e.g. mixing, then we can
find that the chaotic and unpredictable behavior
of some observables 
coexists with a regular and predictable behavior
of densities (so states).
The time evolution of a quantum system is described
either by Schr\"odinger equations (quantum counterpart
of the Liouville picture) or by the Heisenberg equations
(quantum counterpart of the Hamilton's picture).
We shall indicate  how these scheme may be used to
explain the just mentioned coexistence
now in the quantum mechanical description.
In other words, we shall show that a kind of unpredictability can also
exist for quantum maps. To study this phenomenon
we shall use quantum Lyapunov exponents.
To make this point comprehensible let us point out
that we shall study the stability of dynamical maps without any analysis 
of the integrability of the corresponding equations of motion.

In Sections II and III we review the structure
of classical mechanics and basic properties
of (classical) chaotic systems respectively.
This material is mostly well known; however,
we emphasize the role of the picture in the description
of the time development of systems.
Section IV deals with the question of 
(non)linear lifting of dynamical maps.
The objective here is to show the possibility
of nonlinear lifting of such maps as well as
its relation to certain algebraic structure.
Section V provides a detailed exposition
of quantum counterparts of the previously reviewed
characteristic features of classical dynamical systems.
Section VI contains the definition of quantum
characteristic exponents and their basic properties
while Section VII indicates how
these exponents may be used to study stability
of quantum dynamical systems.
We close, Section VIII, with a brief discussion
on the obtained new results.
For the reader's convenience we include the Appendix
with some basic "$C^*$-algebraic vocabulary".

\section{The structure of classical mechanics.}
We start with a review of the essential facts of classical mechanics.
Let $\Gamma \equiv \{q_1,...,q_n,p_1,...,p_n\} $ be  
the phase space of some classical system.
The Newtonian systems with (local) forces
derivable from a potential form the important
and large class of dynamical systems.
The dynamics of this class is described by
the Hamilton's equations:
\begin{equation}
\label{1}
\dot{q_i} = \frac{\partial H}{\partial p_i},
\qquad \; \dot{p_i} = - \frac{\partial H}{\partial q_i}
\qquad i=1,...,n
\end{equation}
where $q_i$,$p_i$ denote generalized coordinate and
momentum respectively, and $H=H(q_1,...,q_n,p_1,...,p_n)$
is the Hamiltonian of the system.
Let $\Omega \subset \Gamma$ be a subset.
We denote by $\A$ the set of all complex-valued smooth
functions on $\Omega$. $\A$ is the set of classical observables
of the considered system.
The Hamilton's equations (\ref{1}) lead
to one parameter group of maps of the phase space
into itself

\begin{equation}
\label{2}
G_t: \Gamma \to \Gamma
\end{equation}
where $G_t\mu $, $\mu \in \Gamma$, is the solution
of the Hamilton's equations with the initial condition 
$G_t\mu|_{t=0} = \mu_0$.
The transformations $G_t$ can be lifted on the level
of observables in the following way

\begin{equation}
U_t: \A \to \A
\end{equation}
where

\begin{equation}
\label{3}
U_tf(\mu) \equiv f_t(\mu) = f(G_t \mu)
\end{equation}
for any $\mu \in \Gamma$.
Obviously, $U_t$ is the one parameter family
of linear maps on $\A$. The prescription given by $(\ref{3})$ is sometimes
called the Koopman's construction.
The next observation is

\begin{equation}
\label{4}
\frac{\partial f_t(\mu)}{\partial t}
=  \sum_{i} \frac{\pa f_t}{\pa q_i} \frac{\pa H}{\pa p_i} 
- \frac{\pa f_t}{\pa p_i} \frac{\pa H}{\pa q_i}
\equiv \{ H, f_t \}
\end{equation}
Here $\{\cdot, \cdot \}$ denotes the Poisson bracket
which can be also considered as a derivation on the abelian algebra 
$\A$ of classical observables (cf. Appendix).
In other words, if we equip $\A$ with the structure of abelian 
$C^*$-algebra then the Poisson bracket determines one parameter
group of automorphism of $\A$ and this group
of automorphism represents the Hamilton's 
evolution of observables.
Another basic concept in the description of classical systems 
is that of a state.
Namely, a state of a system is described by a (probability)
distribution $\varrho$ on $\Gamma$. In particular, a pure state is 
given by

\begin{equation}
\label{5}
\varrho(q_1,...,q_n,p_1,...,p_n)
= \delta(q_1 - q^0_1)\cdot ... \cdot \delta(q_n - q_n^0)
\cdot \delta(p_1 - p_1^0)\cdot ... \cdot \delta(p_n - p_n^0)
\end{equation}
where $\delta(x)$ is the Dirac's $\delta$-function, 
$(q_1,...,q_n,p_1,...,p_n)$ is an arbitrary point
in $\Gamma$ while the considered fixed point in $\Gamma$
is denoted by $(q_1^0,...,q_n^0,p_1^0,...,p_n^0)$.
We remind the reader that evolution in classical mechanics can be 
given in one of two pictures. The one just described 
is so called the Hamilton's
picture, i.e. time evolution of an observable
is given by (\ref{4}) while a state is time independent:
$\frac{d\varrho}{dt} =0$.
The second one, so called the Liouville picture
determines the evolution of a state, i.e. 
the time development of a distribution
$\varrho$. The relevant equations of motion are
\begin{equation}
\label{6}
\frac{df}{dt} = 0, \qquad \qquad \frac{d \varrho_t}{dt}
=- \{H, \varrho_t \}.
\end{equation}
Obviously, on the set
$\{ \varrho - \rm{ an \; arbitrary \; distribution} \} \equiv \St$  
of distributions (states) 
one can define maps $V_t$ (in the same way as in (\ref{3}))
\begin{equation}
V_t \varrho \equiv \varrho_t.
\end{equation}
Clearly, $V_t (\alpha \varrho_1 + (1 - \alpha)
\varrho_2) = \alpha V_t (\varrho_1) + (1 - \alpha)
V_t(\varrho_2)$ for any $\alpha \in (0,1)$.
We remind that these pictures are equivalent in the sense
\begin{equation}
\label{7}
<f_t>_{\varrho} \equiv \int_{\Gamma} f_t(\mu) \varrho(\mu) d\mu
= \int_{\Gamma} f(\mu) \varrho_t(\mu) d\mu
\equiv <f>_{\varrho_t}.
\end{equation}
Thus, the following picture is emerging.
The time evolution in classical mechanics can be described by either
linear $(U_t)$ or convex $(V_t)$ maps. Therefore, there does not exist
a room for nonlinearity which is a necessary condition
for a chaotic behavior.
But we remind (cf. Introduction) that there are many examples 
of concrete models with chaotic behavior. To explain this apparent 
contradiction let us make an important observation.
Although the Hamilton's and Liouville pictures are equivalent
in the sense given by (\ref{7}) the Hamilton's picture
offers larger possibilities. Namely, one can pick up
a coordinate (or momentum)
as an observable. Then
the equations of motion given by (\ref{4}) are reducing to
that of the form (\ref{1}). It is important
to realize that equations (\ref{1}) are on the phase space $\Gamma$
(or on its subset) and for some choice of the Hamiltonian
$H(q,p)$ one can obtain
a system of nonlinear differential equations.
To understand this observation within
the above presented structure of classical mechanics it is 
enough to note that to get the mentioned 
nonlinear equations {\it we did not use
the linear lifting, i.e. the Koopman's 
construction.}
{\it Therefore, for a particular choice of
(classical) observables and the Hamiltonian, there
exists a chance to get a nonlinearity, so also
to get trajectories with unpredictable behavior.}
There is another possibility to get a nonlinearity
for classical systems. Namely, one can ask for a 
non-linear version of Koopman's construction.
However, this is a more subtle point and we will discuss
this question later on.

\section{Chaotic (classical) systems.}
Let us consider a system of differential equations given by (\ref{1})
where for simplicity we put $n=1$.
Suppose $\Rn \ni t \mapsto x^i(t) \equiv (q^i(t), p^i(t))
\in \Gamma$, $i=1,2$
is a solution of (\ref{1}) with initial conditions
$x^i(0) = (q^i(0), p^i(0))$, $i=1,2$.  
Put $\Delta x(t) \equiv dist(x^1(t),x^2(t))$.
Then one can ask the following question: what is behavior
of the function $\Rn \in t \mapsto \Delta x(t)$?
We are interested in the class of systems such that
the growth of $\Delta x(t)$ is of the exponential type.
We remind that the exponential growth is measured
by Lyapunov exponents which can be defined as
the following limits:

i) for a discrete dynamical system:

\begin{equation}
\label{8}
\lambda^{\rm cl}(x,y) = \lim\limits_{n\to\infty}\,{1\over n}\,\log 
\vert D_x{\tau}^n (y) \vert 
\end{equation}
where $D_x{\tau}^n(y)$  denote the directional derivatives of ${\tau}$ 
composed with itself $n$ times at a point $x$ in a direction
$y$ (cf. \cite{ER}). 

ii) for a continuous dynamical system:

\begin{equation}
\label{8 a}
\lambda^{\rm cl}(x,y) = \lim\limits_{t\to\infty}\,{1\over t}\,\log 
\vert D_x{\tau}_t (y) \vert 
\end{equation}
where $D_x{\tau}_t(y)$  denotes, as above, the 
corresponding directional derivative.  

\begin{rem}
\label{Lap}
If we examine a time evolution $\tau_t$ of a point $x$
depending on a parameter $\upsilon$, then
\begin{equation}
\label{9}
lim_{t \to \infty} \frac{1}{t} log | \frac{d}{d \upsilon}
\tau_t(x(\upsilon))|
\end{equation}
also can be considered as a measure of 
sensitivity of evolution with respect to initial conditions
(determined by $\upsilon$)
provided that the limit in (\ref{9}) exists. This limit
can be also called Lyapunov exponent.
\end{rem}
Both limits measure the average rate of 
exponential
growth of separation of orbits which at time zero differ by a small vector. 
This property is used as a measure
of the sensitivity of dynamical system with respect to
initial conditions.
\begin{defn}
\label{Reg}
A dynamical system described by differential equations
of the type (\ref{1}) is called regular (irregular) if $\lambda^{cl} \leq 0$
($\lambda^{cl} >0$ respectively).
\end{defn}
Lyapunov exponents are closely related with
Kolmogorov-Sinai entropy  $\E_{K-S}$ which is another
key quantity for analysis of classical dynamical systems.
 $\E_{K-S}$ measures mixing properties of a system and its relation
to Lyapunov exponents for smooth enough systems is given by
(cf. \cite{Miliun}, \cite{Pesin1}, \cite{Pesin})
\begin{equation}
\label{arka}
\E_{K-S} = \sum_{k; \lambda_k \geq 0} \lambda_k
\end{equation}
Thus, for a large class of dynamical systems, 
a positivity of Kolmogorov-Sinai entropy indicates
the existence of positive Lyapunov exponents.
We recall that for more general class of dynamical
systems (that is satisfying weaker smooth conditions)
one should replace equality in (\ref{arka}) by an inequality
(see {\cite{LedraYoung}).
\begin{examp}
\label{hip}
Let us consider a dynamical system determined by
the following differential equation
\begin{equation}
\label{11}
\frac{d^2}{dt^2} x = \kappa^2 x, \qquad
\kappa \geq 0.
\end{equation}
The solution of (\ref{11}) can be written as
$x(t) = x_1 cosh(\kappa t) + x_2 sinh(\kappa t)$
where $x_1,\; x_2$ are constants describing the initial conditions.
Put, for simplicity, $x_2 =0$ and assume that we have two 
initial conditions $x_1^I$ and $x_1^{II}$ such that
$|x_1^I - x_1^{II}| = \epsilon$. The corresponding solutions will
be denoted by $x^I(t)$ and $x^{II}(t)$ respectively.
We note
\begin{equation}
\Delta x(0) \equiv |x^I(0) - x^{II}(0)| = 
|x_1^I - x_1^{II}| = \epsilon
\end{equation}
while
\begin{equation}
\Delta x(t) = \epsilon \; cosh(\kappa t) \approx_{t \to \infty}
\frac{\epsilon}{2}\; e^{\kappa t}
\end{equation}
Thus, the considered system exhibits 
the exponential growth of separation of trajectories,
i.e. a sensitivity with
respect to initial conditions. However, it is important
to realize that it is difficult to consider
this system as a chaotic one.
\end{examp}
This example clearly indicates that a definition
of chaotic system presents a delicate problem. In particular, 
in the above example, 
we have gotten a sensitivity with respect to initial conditions
since we allow trajectories to spread to infinity.
This motivates
\begin{defn}
\label{chaos}
A (classical) dynamical system exhibits
a chaotic behavior if 
\begin{itemize}
\item it is defined on a compact
subset of the phase space (or the algebra
of observables is defined on a compact subset of the 
phase space),
\item it has an irregular motion with
some ergodic properties.
\end{itemize}
\end{defn}
Let us comment on this definition. Firstly,
the assumption of compactness excludes a possibility 
of a system with trajectories going to infinity.
In other words, an irregular evolution should have a kind of repeatibility.
Let us emphasize that all computer simulations
of chaotic systems have this feature.
Next, we wish to briefly comment on the assumption
of ergodicity. In real dynamical systems
there are coexisting regions of chaotic behavior 
with regions of regular motion.
So we need a kind of ergodicity
to get a system with overwhelming
region of irregular motion.

We want to close this Section with the two remarks
on the just given definition of chaotic system.
Firstly, we want to point out that the discussed definition
is not the only one possible.
On the contrary, there are many different approaches to this topic with
various definitions of chaos.
Even for {\it semigroups of linear operators on a Banach
space, there is an attempt to introduce
a concept of chaos} (see the definition
given by Deveney \cite{Den},  see also \cite{DWW}).
However, we strongly believe that for real mechanical
systems the presented definition is the proper
choice as it reflects our intuition on sensitivity
of dynamics. Secondly, we want to stress again that the
Liouville picture seems to be not the best
choice for doing a study of chaotic systems.
Clearly, one can examine the basic features of the
Peron-Frobenius operator.
Nevertheless, we think that
such the work can only give some insight for
"shadows" of the real chaotic behavior.

\section{Is The Koopman's construction the only
one possible?}
In the previous Section it was argued that in the study of
chaotic systems we should restrict the class of dynamical 
systems to that defined on a compact subset of the phase 
space $\Gamma$. This implies that the algebra of observables
associated with such the system is the set of all complex
valued continuous functions defined on $\Omega$
where $\Omega$ is a compact subset of $\Gamma$.
Furthermore, equations of motion for the system 
- the Hamilton's equations (\ref{1})
should be restricted to $\Omega$. But one can pose a question:
what is the proper choice for $\Omega$.
If we consider a conservative system (such systems have
the dynamics given by the Hamilton's equations) the most natural
choice is the isoenergetic surface. 
So, in the remainder of this section we assume that $\Omega$
is a compact isoenergetic surface of some dynamical system.
Consequently, $\Omega$ will be also considered as a compact topological space.
We denote by
$\com$ the Banach space of all $\Cn$-valued
continuous functions on $\Omega$. 
$\com$ with pointwise defined multiplication, complex conjugation
and the supremum norm is an abelian unital $C^*$-algebra.
By the way, we stress that {\it the compactness of $\Omega$
is ``translated'' in the algebraic language by
the property of $\com$ to be a unital algebra.}
But now we have the following problem:
$\com$ has the very rich algebraic structure, in particular
one can write a square. So the concept of nonlinear
map on $\com$ is perfectly well understood (cf. Example 2,
subsection VC).
On the contrary, the situation on $\Omega$ is much
less clear. To see the problem let us
consider, as an example, one dimensional
harmonic oscillator. Then $\Omega$ in this case is just 
the ellipse. So, a natural question is emerging:
{\it Is it possible, starting from $\Omega$, to define 
such an algebraic structure $\cW $
which is compatible with that of $\com$? }
In the just mentioned example of a harmonic 
oscillator, the ellipse can be naturally equipped
only with topological and measure structures
and these structures are quite enough for
the standard Koopman's construction.
However, the ellipse has not an algebraic structure
and therefore it is impossible to study
a nonlinear lifting in this setting.
To clarify this question let us emphasize that, here, 
the nonlinear lifting is understand as the
procedure leading from a nonlinear map
defined on $\Omega \subset \Gamma$ to
a nonlinear map defined on the set of all observables.
To solve the problem we propose to use the 
construction which was called
``the  nonlinear version of the Koopman's
construction'' (see \cite{MM}).
To describe the solution of our problem
we need some preliminaries.
Let us recall that a pure state $\vp$ on 
an abelian $C^*$-algebra $\A$ (called also a
character) is a linear, continuous, positive and normalized functional
on $\A$ such that $\vp(ab)=\vp(a)\vp(b)$
for $a,b \in \A$ (cf. Appendix). 
Denote the set of all characters by $\cP$. 
Let us add that by Gelfand-Naimark
theorem we can identify an abelian $C^*$-algebra $\A \equiv \com$ with
the algebra of all continuous functions on $\cP$ where
$\cP$ is equipped with the pointwise convergent topology.
Moreover, one can establish one-to-one
correspondence between $\Omega \ni x \mapsto \vp \in \cP$.
So to keep notations of the previous sections we shall
identify $\om$ with $\cP$.
Let $\cW_\om $ denote the algebra of continuous functions on 
$\com$ generated by the set $\cP (= \Omega)$ of all characters of $\com$.
Elements of $\cW_\om$ are of the form 
$\vartheta = W(\vp_1,...,\vp_n)$, where $n\in \Nn$, $W$ is 
a polynomial of $n$ commuting variables with
complex coefficients, and $\vp_1,...,\vp_n\in\cP$.
Using the correspondence between the set $\om$ and
the set of all characters (cf. \cite{MM}, see also
\cite{KadRin} for details of the theory of operator
algebras)
one has the following inclusion
\begin{equation}
\om\subset \cW_\om .
\end{equation}
Now, let us remind that the Gelfand transform is defined as a map
\begin{equation}
\widehat{ }:\A\to C(\cP)
\end{equation}
which to every $a\in\A$ assigns a function
\begin{equation}
\label{transformata}
\widehat{a}(\vp) \stackrel{\rm def}{=} \vp (a) .
\end{equation}
where $\vp$ is a character. Consequently, 
$C(\cP)$ can be considered as the image of $\A$ under the Gelfand 
transform.

Let $a\in\A$. We define the following complex valued function $\tilde{a}$
on $\cW_\om$:
\begin{equation}
\label{extension}
\tilde{a}(W(\vp_1,...,\vp_n)) =W(\vp_1(a),\ldots ,\vp_n(a)).
\end{equation}
It is easy to show that $\tilde{a}$ is in fact an extension of
$\widehat{a}$ over $\cW_\om$. Namely, in the case $n=1$ in 
(\ref{extension}) one has
\begin{equation}
\label{podstawowa}
\tilde{a}(W(\vp)) = W(\vp (a)) = \vp (W(a)) =\widehat{W(a)}(\vp ).
\end{equation}
Here the second equality follows from multiplicativity 
of the character $\vp$. Further,
we shall consider dynamical maps on $\cW_\cP$ which are of the following
form
\begin{equation}
\label{teta}
\theta (W(\vp_1,\ldots,\vp_n))=(T(W))(\tau(\vp_1),\ldots,\tau(\vp_n)),
\end{equation}
where $T$ is a polynomial of one variable, and $\tau :\cP\to\cP$ is
a continuous map. Let us denote the set of all maps in this form
by $\Theta$.
Now we are in position to give (cf. \cite{MM})
\begin{theo} 
\label{twierdzenie}
For an arbitrary map $\theta\in\Theta$
there exists, in general nonlinear, the well defined map
\begin{equation}
C(\cP) \ni\widehat{a}\mapsto U_\theta\widehat{a}\in C(\cP)
\end{equation}
and $\theta$ is uniquely determined by $U_\theta$
where
\begin{equation}
U_\theta = U_\tau \circ T. 
\end{equation}
$U_\tau$ is the operator defined by
$(U_{\tau}f)(x)=f(\tau (x))$.
\end{theo}
Thus, we got the affirmative answer to the question posed in the
beginning of this Section. There is a possibility
to introduce the algebraic structure 
$\cW_\om$ in such a way that the multiplication
in this structure is obtained from that of $\A$.
Therefore, nonlinearity of a map on $\cW_\om$ is 
compatible with the nonlinearity of the map on $\A$.
This gives a possibility of nonlinear lifting
of a map on $\cW_\om$ to a map on $\A$.

To make the basic idea of the presented construction
more clear let us again consider
the example of one dimensional harmonic oscillator.
Then, the set $\om$ is equal to the ellipse.
$\cW_{\om}$ can be considered as ``an algebraic reconstruction''
of the phase space $\Gamma$ (from the singled out
subset $\Omega$) in such a way that
its multiplication is compatible with
the natural multiplication of the algebra
of (classical) observables $C(\om)$.
The main difficulty in carrying out this
reconstruction is that there is still one question 
unanswered. {\it Namely, we do not know a direct relation
between $\Gamma (\supseteq \Omega)$ and $\cW_{\om}$}.
On the other hand, this nonlinear version of Koopman's
construction suggests rather strikingly the explanation why
among the well known examples of chaotic systems
the periodically perturbed ones play a prominent role.
Such the systems, in general, are not conservative ones.
Let us consider 
the one dimensional kicked harmonic oscillator;
a nearly paradigm of chaotic system.
Let $\Omega$ be again the isoenergetic surface of the 
harmonic oscillator. To build $\cW_{\Omega}$ we should construct
over each point of $\Omega$ a ``fibre'' which is 
a free ring of one variable.
Then one can suppose that  
the free motion is determined by a map $\tau$ 
defined on the isoenergetic surface, i.e.,
$\tau : \Omega \to \Omega$.
The Theorem \ref{twierdzenie} says that $\tau$ can be
linearly lifted to a map on $\A$. Then a kick (perturbation)
acts on the system. Within this picture, one may say that
a kick corresponds to a motion along a fiber.
Consequently, a kick is reflected on the set of observables as a nonlinear
map (see Theorem \ref{twierdzenie}).
Thus, we have got the explanation saying that
kicked systems can be related with the ``algebraic''
nonlinearity. We remind, a nonlinearity is the necessary condition
for a chaotic behavior.

Having clarified the basic concepts for classical dynamical
system we want to pass to the basic subject of the paper, namely to 
the problem of existence and description of quantum chaos.

\section{Quantum mechanics }
It was pointed out in Section II that the basic
concepts for a description of a classical
dynamical system are: algebra of observables, states, 
and prescription for the equation(s)
of motion which can be formulated in one of two pictures 
(equivalent in the sense of equality (\ref{7})).
To quantize the (classical) scheme it is enough to change
the realization of the algebra of observables with
the relevant modifications of description of
state and evolution. 
Therefore, 
we replace the algebra $C(\om)$ (where $\om$ is a compact
subset of $\Gamma$) by a unital (non-abelian)
$C^*$-algebra $\A$. We repeat: {\it the unitality of the algebra
is the algebraic translation of the topological
compactness property!}
Then, the states are linear positive, normalized
(so continuous) functionals on $\A$. The time evolution
is described by quantum counterparts of the Liouville and Hamilton's
pictures, i.e. by the Sch\"odinger
and the Heisenberg picture respectively.
Our objective in this Section is to convince
the reader that the Heisenberg picture is the suitable framework
for investigations of quantum chaotic problems.

\subsection{The Schr\"odinger picture}
\subsubsection{Linear Schr\"odinger equation} 
This is the quantum counterpart of the Liouville picture for a pure state.
The basic equation of motion is
\begin{equation}
\label{schrod}
i\frac{h}{2\pi} \frac{d\Psi(t)}{dt} \equiv
i \hbar \frac{d\Psi(t)}{dt}  = H \Psi(t)
\end{equation}
where $\Psi(t)$ denotes the wave function. Its solutions are 
of the form
\begin{equation}
\Psi(t) = e^{- \frac{i}{\hbar} H t} \Psi(0)
\end{equation}
Thus
\begin{equation}
\label{V1}
|| \Psi(t) - \Psi^{\prime}(t) || =
|| \Psi(0) - \Psi^{\prime}(0) ||
\end{equation}
We observe that the equality (\ref{V1}) implies that
{\it there is not any growth of separation
of orbits which at time zero differ by a small vector!}
Thus there does not exist 
a room for a sensitivity on initial conditions. 
Consequently,
there is no chance for positive Lyapunov exponents.
{\it We want to emphasize that exactly the same situation
we have for a classical dynamical system with
dynamics given in the Liouville picture.}
Not taking into account the last remark, the long
debate on existence of quantum chaos has led to
a conjecture that quantum mechanics suppresses chaotic
behavior in such a way that there is only a room
for some ``shadows'' of chaos.
Adopting this point of view one could only study
special features of a wavefunction which describes
quantum counterpart of classically chaotic system
and to investigate some properties of the correspondence
between classically chaotic system and its quantum
counterpart.
In other words we have the quantum analogy of the
search of chaotic signatures within
the Liouville picture (cf. the last paragraph
of Section III).

Let us briefly discuss a typical result in this direction.
The most famous proposition concerning a characterization
of special properties of quantized classically chaotic systems
is given by the following
conjecture: There is a correspondence
between classical (regular or irregular) dynamical system
and the energy level statistics of the Hamiltonian
of the quantum counterpart of the considered system.
Namely,
\begin{eqnarray}
regular \; (classical) \; system  \quad & 
\leftrightarrow & \quad Poisson \; distribution \\
irregular \; (classical) \; system \quad & 
\leftrightarrow & \quad Wigner-Dyson \; distribution \\
\nonumber
\end{eqnarray}                                                              
where the Poisson distribution is given by
$P(\Delta E) = a e^{-b \Delta E}$, $a, b$ are some positive
constants while the Wigner-Dyson distribution
is given by $P(\Delta E) = A |\Delta E|^{\alpha}
e^{- \beta (\Delta E)^2}$, with $A, \alpha, \beta$
some positive constants.
However, recently (see \cite{Craham}) it was shown that there are
infinitely many counterexamples to the above conjecture.
Thus, although there are some computer experiments
verifying this conjecture for
some models it is difficult to say
that there is a well defined correspondence 
between energy level statistic and the nature of evolution.

As a matter of fact the same picture is emerging from
the analysis based on the other apparently 
basic features of the wave function
of quantized (classically) chaotic system.
To specify them let us list: scars, revivals, localization
of wave function, level repulsion leading to regions
of complicated avoiding crossing.
To the best of our knowledge, there does not exist any
rigorous proof of a correspondence between
such features and quantum chaotic systems.
Therefore, looking for an answer to the question
whether there is ``true'' chaos in quantum mechanics
we drop this way of investigation of
the posed problem.

\subsubsection{Non-linear Schr\"odinger equation}
Let us consider the Schr\"odinger type
equation (in general, nonlinear)
\begin{equation}
i\hbar { d \psi \over{ dt}} = H_0 \psi + V(\psi) 
\label{S1}
\end{equation}
where $\psi \in \cal H$, $\cal H$ is a separable Hilbert space,
$H_0$ is the free Hamiltonian (yielding a linear evolution), and 
$V(\psi)$ is a (nonlinear ) potential which is defined
below. In other words, we begin with a field-theoretic model
where the states at a given time are represented by vectors
in $\cal H$ while the nonlinearity enters through their
time-dependence, i.e. the nonlinearity is described by (\ref{S1}).
To give the definition of $V(\cdot)$ let us 
denote by $\{ e_k \} $ a basis (independent of t) in $\cal H$.
Define 
\begin{equation}
{\cal H} \ni \psi \mapsto U(\psi) = \{ <e_k, \psi> \} \in \ell_2
\label{S2}
\end{equation}
($\ell_2$ is the Hilbert space of sequences $\{ \alpha_i \}$, $\alpha_i$
a complex number, such that $\sum_i \vert \alpha_i \vert^2 < \infty$.)
Denote by $U$ the unitary map from $\cal H$ onto $\ell_2$.
We assume that a function $v( \{ \alpha_i \} ) \buildrel
\rm def \over = 
\{ v(\alpha_i) \}$ satisfies the following
condition
\begin{equation}
\sum_i \vert v(\alpha_i) \vert^2 < \infty
\quad {\rm provided \quad that} \quad \sum_i \vert \alpha_i \vert^2 < \infty
\label{S3}
\end{equation}
The nonlinear term $V(\psi)$ in (\ref{S1}) is defined by
\begin{equation}
V(\psi) = U^{-1} v(U(\psi)) 
\label{S4}
\end{equation}
Let us remark that if one takes another basis, say $\{ f_i \}$,
in $\cal H$ then (\ref{S2}), (\ref{S3}), and (\ref{S4}) give, in general, 
another nonlinear operator
$V^{\prime}(\psi)$ on $\cal H$. However, one has
\begin{equation}
W^{-1}V^{\prime}(W \psi) = V(\psi)
\label{S5}
\end{equation}
where $W$ is the unique unitary operator on $\cal H$ such that
$W( \{  e_k \}) = \{ f_k \}$. Thus the operator $V(\psi)$ 
is defined in a "covariant" way.

As examples of nonlinear potentials leading
to nonlinear type Schr\"odinger evolution (\ref{S1}) we can mention
$V_1(\psi)= \overline{\psi} \psi \psi $
(its Heisenberg counterpart has applications to molecular
dynamics and nonlinear optics, \cite{CJLP} see also
 \cite{Mil}, \cite{MH}) or
$V_2(\psi) = log(\psi) $ for $\psi \in {\cal H}_0 \subset
{\cal H}$ where the domain ${\cal H}_0$ 
of the (unbounded) operation $log(\cdot)$ 
is determined by the condition (\ref{S3}), 
(such evolution describes frictional
effects in dissipative systems, \cite{K}, \cite{SChH}). 
Other examples of nonlinear Schr\"odinger equations can be
found in (\cite{DG}, \cite{BBM}, \cite{Kib}).

However, {\it one may ask whether the equation (\ref{S1})
is of the fundamental nature or it is a kind of
approximation}. Although there are some arguments
that nonlinear Schr\"odinger type equations could be
fundamental ones (see e.g. \cite{DG},
see also the argument given in Subsection VB), 
this question is still unanswered.
Namely, one can also consider this type of equation
as a result of an assumption that a part of a larger quantum
system was regarded as ``heat bath''.
The lack of convincing arguments can be also related to
the question of the nature of frictional effects.
If one admits the conjecture that the Newton's
dynamics is more general one than that of the Hamilton's
the answer to the just raised question could be affirmative,
i.e. one can expect to find ``genuine''
nonlinear quantum maps. 
Let us add that, in such the case, even on the level of classical mechanics
the Liouville picture based on the Hamilton's
dynamics would be not the most general
prescription for the evolution in the sense that nonlinear
corrections would be necessary ingredient of the equations of motion.

Nevertheless, having a nonlinear Schr\"odinger
equation one can ask for a possibility of 
chaotic behavior. To answer this question
one can proceed as follows.
We begin with rewriting (\ref{S1}) in terms of Fourier coefficients, i.e.
using the basis $\{ e_k \equiv \vert k> \}$ 
in $\cal H$ and the formula (\ref{S1}) we arrive to
\begin{equation}
 i \hbar {d \over dt} \sum_k <k \vert \psi>  \vert k> = 
\sum_k <k \vert H_0 \psi> \vert k> +
\sum_k <k \vert V(\psi)> \vert k>.
\label{S6}
\end{equation}
Now, just for simplicity, let us put $H_0 = 0$. 
Moreover, to give as simple as possible
arguments we take a concrete form
of the nonlinear potential.
Thus, in
the definition of $V(\cdot)$ let us put 
$v(\alpha) = \kappa \overline{\alpha} \sum_n \delta(t - nt_0)
+ {\chi \over 2} \overline{\alpha} \alpha \alpha$ where
$\kappa$ and $\chi$ are positive constants, $t_0$ is a fixed 
period of time. Such potential was designed to describe
the nonlinear evolution in the Kerr medium with kicks where the kicks
are defined by the first term in the formula for $v$, i.e.
$\kappa \overline{\alpha} \sum_n \delta(t - nt_0)$ 
(for details see \cite{Mil}, \cite{MH}).
Hence
\begin{equation}
i \hbar {d \over dt} z_k = \kappa \overline{z_k} \sum_n \delta(t - nt_0)
 + {\chi \over 2} \overline{ z_k} z_k z_k \quad k=1,2,...
\label{S7}
\end{equation}
where $z_k = <k \vert \psi>$.  
Let us note that owing to the properties of Dirac's $\delta$, 
(\ref{S7}) can be considered as a system of two equations; the first one
determines the dynamics in the period between kicks while
the second one gives the parametric process describing kicks in the 
time $t_n = nt_0$, $n=1,2,...$ (cf. \cite{MH}).
Clearly, (\ref{S7}) is the (standard) nonlinear differential equation
and an application of classical methods leads to solutions of (\ref{S7}) 
which  admit
a positive Lyapunov exponent (cf. \cite{Mil}, \cite{MH}). 
In other words (\ref{S1}) leads to an example
of nonlinear Schr\"odinger type equations with sensitive solutions
with respect to initial conditions where the Lyapunov exponent, $\lambda^q_s$,
is defined as $\lambda^q_s = \lim_{t \to \infty} {1 \over t}
log{ \Vert \delta \psi(t) \Vert \over \Vert \delta \psi(0) \Vert}$
and $\Vert \psi \Vert$ is expressed in the Fourier coefficient terms
(cf. \cite{CJLP}). So we conclude that within the nonlinear
type of Schr\"odinger evolution there is a room for a
chaotic behavior.

\subsubsection{Density matrix formalism}

The genuine quantum counterpart of the Liouville picture
is that given in terms of density matrices. So, 
let us consider the quantum
Liouville equation, called the von Neumann equation,
which is of the form
\begin{equation}
\label{D1}
\dot{\varrho} = - \frac{i}{\hbar} [H, \varrho]
\end{equation}
where $H$ stands for the Hamiltonian of the considered system,
$\varrho$ is a density matrix while $[\cdot, \cdot]$
denotes the commutator.
 Clearly (\ref{D1}) is a linear differential
equation and one can ask for nonlinear
version of this equation. 
We recall that  nonlinear equations for time evolution
of density matrix
are known and used in physics (cf. \cite{ALM}, \cite{Bel}, \cite{Spohn}).
Furthermore, similarly as for the Schr\"odinger equation, the question 
concerning the nature of the equation (\ref{D1})
as well as its nonlinear corrections can be posed.
Basically one has the same answers.
Therefore we will not discuss this point here;
let us pass to an illustrative example instead.
Let us consider a dynamical system with dynamics given by Hartree -
type evolution equation (cf. \cite{AM}):
\begin{equation}
\label{D2}
\frac{d}{dt} {\rho }_t = - \frac{i}{\hbar} [H({\rho }_t ), {\rho }_t ] 
\end{equation}
\begin{equation}
\label{D3}
{\rho }_t {\vert }_{t=0} = {\rho }_o ,
\end{equation}
where $H({\rho }) = [Tr ({\rho }Q)]Q$ with $Q = Q^*$,
and $\varrho$ is a density matrix on a finite dimensional
Hilbert space $\cal H$. We recall that this type of
equation can be obtained in the mean field limit for interacting
quantum systems. The explicit solution of (\ref{D2}) is given by
\begin{equation}
\label{D4} 
{\Phi }_t ({\rho }_o ) = \exp {[- \frac{i}{\hbar} Tr(Q{\rho }_o )Qt]} 
{\rho}_o \exp {[\frac{i}{\hbar} Tr(Q{\rho }_o )Qt]}. 
\end{equation}
Clearly, the solution $\Phi(\cdot)$ of (\ref{D2}) 
is a nonlinear quantum map.

To summarize, within the quantum Liouville picture 
there are examples of nonlinear evolution
equations. Consequently, this can be taken as
a motivations for our investigations
of stability of dynamics determined by such the equations.

\subsection{The Heisenberg picture}
We recall that this is the quantum counterpart 
of the Hamilton's picture (the picture from the classical mechanics,
cf. Section II).
The fundamental equation of motion is of the form
\begin{equation}
\label{H1}
\frac{d}{dt} A = \frac{i}{\hbar} [H,A]
\end{equation}
where $A$ is an operator in
the algebra of observables,
while $H$ stands for the Hamiltonian of a system.
The solution of (\ref{H1}) can be written in the form
\begin{equation}
\label{H2}
A \mapsto e^{\frac{i}{\hbar} Ht}Ae^{- \frac{i}{\hbar} H t}
\end{equation}
Our first remark is that the Heisenberg
formula for equations of motion is in
harmonious relation with the very rich algebraic $C^*$ 
structure of the matrix formulation of quantum mechanics.
On the contrary, the Schr\"odinger picture
does not have this property: a Hilbert
space has only a linear space structure
while the set of density matrices for a real
physical system is a Banach $^*$-algebra. 
There is only one exception. Namely, 
for finite dimensional models of quantum mechanics,
the set of trace class operators is equal to
the set of all bounded operators. Therefore,
for such particular models one can also use the $C^*$ algebraic
technique for an analysis of evolution
of density matrices.
The next remark is related to a possibility of getting 
nonlinear operator
equations. To be more specific, 
let $H$ be a Hamiltonian
of a physical system such that $H=H(A,B)$ is a function 
of several noncommuting dynamical variables. 
Obviously, in general, the definition of a function
of two (or more) noncommuting variables is not
a clear stuff. However, in Quantum Mechanics there are many examples
of Hamiltonians of this type which
have arisen as a result of the quantization 
procedure. So, here, we can put for the concrete $H(A,B)$ 
the relevant Ansatz.
Then, the solution
of the Heisenberg equation of motion for $A$ (the equation needs 
not be linear in $A$)
\begin{equation}
A\mapsto e^{\frac{i}{\hbar} H(A,B)t}A
e^{-\frac{i}{\hbar} H(A,B)t}
\label{H3}
\end{equation}
does not have to be the linear function (in $A$).
Clearly, we have not such a possibility for 
the Schr\"odinger evolution of pure
states.

Let us consider the question of nonlinear dynamical
maps in the Heisenberg picture in detail.
As it was mentioned at the beginning of this Section
the Heisenberg picture is quantum counterpart of the Hamilton's picture.
Let us remind (cf. Section II) that the Hamilton's
equations also lead to a one parameter group
of linear maps on the algebra $\A$ of observables
where $\A \equiv \com$. In particular, the Poisson
bracket is nothing else that the derivation on $\A$
(cf. Appendix).
However, as {\it it was pointed out in remarks
following (\ref{7}) there is a possibility,
for the very special choice of observables,
to study of time evolution
without the linear lifting.}
Namely, the choice of $\{q\}$ and $\{p\}$ can lead to
nonlinear differential equations with solutions
given by nonlinear maps
(even to the solutions
which are sensitive with respect to initial conditions).
Now let us turn to quantum mechanics.
We know, e.g. from quantum optics,
that the creation $a^{\dagger}$ 
and the annihilation $a$ operators can be treated as 
a substitution for $q$ and $p$.
Let us take this point of view
and let us put in (\ref{H2}, \ref{H3})
$A = a^{\dagger}$ , $B=a$.
Then, we can expect to obtain nonlinear dynamical maps
for some choice of Hamiltonians.
Consequently, we can expect (similarly as in classical mechanics)
to find in this way
nonlinear equations and maps with interesting (non)stability
properties. We want to add that another
way of introducing nonlinear maps
in quantum physics will be studied in the next subsection.

We close this subsection with a brief discussion
on the equivalence of Schr\"odinger and Heisenberg pictures.
Similarly, as for the classical mechanics (see (\ref{7}))
we have
\begin{eqnarray}
\label{H4}
<A>_{av}(t) \equiv <A>_{\Psi(t)} = (\Psi(t), A \Psi(t))
= (e^{- \frac{i}{h} H(A,B)t} \Psi(0),A e^{- \frac{i}{h} H(A,B)t}\Psi(0)) \\
=(\Psi(0), e^{\frac{i}{h} H(A,B)t} A e^{- \frac{i}{h} H(A,B)t} \Psi(0))
\equiv <A(t)>_{\Psi(0)} \equiv <A(t)>_{av} \nonumber 
\end{eqnarray}
To get a better understanding of the nature of nonlinearities
described by (\ref{H3}) let us replace $A$ in
\begin{equation}
\label{H5}
A \mapsto  e^{\frac{i}{h} H(A,B)t} A e^{- \frac{i}{h} H(A,B)t}
\end{equation}
by $<A>_{av}$.
This can be done, e.g. by performing some approximations
in averaging procedures (cf. \cite{FE}, \cite{Mil}).
In this way we can arrive to
\begin{equation}
\label{H6}
<A>_{av} \mapsto f(<A>_{av}, t)
\end{equation}
where $f(\cdot)$ is, in general, a nonlinear function.
Thus, we can obtain the ``classical'' nonlinear maps
and we can expect to find examples having sensitiveness
with respect to initial conditions. In fact such examples
were found, e.g. see (\cite{FE}, \cite{Mil})
Obviously, in our discussion of stability properties
of quantum systems we want to avoid the obscure procedure
leading to (\ref{H6}). {\it To this end
we shall introduce quantum Lyapunov exponents} (cf. Section VI).

To finish our discussion on equivalence of both pictures
we should say few words on the relation between the nonlinear
Schr\"odinger equation (\ref{S1}) and the Heisenberg equation
(\ref{H1}). Obviously, the prescription (\ref{H4})
based on bilinear form $(\cdot , \cdot )$ (given
by the scalar product) is not the relevant one. However, 
to pass from a differential equation for a state to
an operator equation involving an observable we can use
"{\it the second quantization rule}" (cf. \cite{BR}, \cite{CJLP},
 see also \cite{Majew1}).
To explain this idea let us restrict ourself to
the model described by equations (\ref{S6}) and (\ref{S7}).
Then, the equation of motion of the annihilation operator $a_i$
(respectively the creation operator $a^*_i$) follows from the equation
of motion (\ref{S7}) in which we replace the Fourier coefficients
$z_i$ and $\overline{z_i}$ (so complex numbers) by
the boson creation and annihilation operators $a_i$ and $a_i^*$
(so operators).
Thus, we have
\begin{equation}
i \hbar {d \over dt} a_j = \kappa  a^*_j  \sum_n \delta(t - nt_0)
 + {\chi \over 2} a_j^* a_j a_j
\label{H7}
\end{equation}
The remark following (\ref{S7}) is also applicable to this case.
Let us observe that (\ref{H7}) follows 
from the Heisenberg equation of motion
\begin{equation}
i \hbar {d \over dt} a_j = [H,a_j]
\label{H8}
\end{equation}
with $H = i {\kappa \over 2} \sum_j [(a_j^*)^2 - (a_j)^2]  
\sum_n \delta(t - nt_0) + 
{\chi \over 4} \sum_j a^*_j a^*_j
a_j a_j$.
Thus, the second quantization methods links nonlinear
Schr\"odinger type equation with the undoubtly fundamental
equation of motion of the Heisenberg form.
Moreover, the latter can lead
to nonlinear operator-valued equations.

To summarize this subsection, we conclude that within the Heisenberg
picture there is a room for quantum
counterparts of ``nonlinear'' type evolution.

\subsection{Are there many nonlinear quantum maps?}

We have pointed in previous Sections that
although nonlinear relations are not sufficient
for chaos of classical systems, some form of nonlinearity
is necessary for chaotic dynamics. 
We observed that the Heisenberg equations of motion
as well as the (nonlinear) Shr\"odinger equations allow
some form of nonlinearities.
Besides we would like to know whether a nonlinearity
can arise from the ``nonlinear''
quantization of nonlinear systems. We remind that in the theory
of classical chaos we frequently study
the dynamical maps alone. This arises the question for
a quantization such the dynamical maps without
referring to Hamiltonian equations.
In other words we are talking about a possibility
of quantization of Newtonian systems.
Working in this direction, we have shown (cf. Section IV)
that there exists the nonlinear version of Koopman's
construction. Here, we indicate how this construction
can be used to argue that the well known class
of nonlinear completely positive maps defined on
a $C^*$-algebra (see Section VI and the Appendix) may be of 
the ``physical'' nature; for another motivation
to study nonlinear completely positive maps
in quantum dynamical systems see \cite{Bel}.

Let again $\A_{abel} \equiv C(\Omega) \equiv C(\cP)$
be an abelian $C^*$-algebra with unit where $\Omega$ is 
a compact (Hausdorff) space (cf. Section IV).
Within the algebraic structure  $\cW_{\om}$
we can consider the map $\vp \mapsto \vp^n$.
We observe
\begin{equation}
\hat{a}(\vp) = \tilde{a}(\vp) 
\mapsto \tilde{a}(\vp^n) = \widehat{(a^n)}(\vp)
\label{H9}
\end{equation}
where $a \in \A_{abel}$. In other words, there exists
the correspondence between the map $\vp \mapsto
\vp^n$, $\vp \in \Omega$, $\vp^n \in \cW_{\om}$ and the map
$a \mapsto a^n$, $a, a^n \in \A_{abel}$.
Consequently, if a subset $\Omega$ of the phase 
space is algebraized properly, 
i.e. it is endowed with the $\cW_{\om}$ structure
we are able to translate nonlinearities of maps
defined on $\cW_{\om}$ to that defined on $\A_{abel}$.
Let us illustrate this point.

\begin{examp}
\label{logistic}
Let us define
\begin{equation}
\cL^d \widehat{f} \equiv r \widehat{f}(1 - \widehat{f}) 
\end{equation}
where $1$ stands for the identity function, $\widehat{f}(\cdot) \in C(\cP)$
while $r$ is a constant.
We have
\begin{eqnarray}
\cL^d\widehat{f}(\vp ) & = & (r\widehat{f}-r\widehat{f}^2)(\vp )=
r\vp (f)-r\vp(f^2)
= r \vp(f) - r \vp^2(f) = \\
 & = & 
\left( r(\vp - \vp^2) \right)(f) 
\equiv (\cL\vp)(f) 
\end{eqnarray}
where $\vp \in \cP$, $ \widehat{f}\in C(\cP)$
and finally $f$ is the element of $\A$ which defines the
Gelfand transform $\widehat{f}$.
Consequently, our example demonstrates rather strikingly that
{\it the logistic map} (one of the most
famous ``chaotic" maps) can be considered both in $\cW_{\cP}$
and $C(\cP)$ terms. Evidently, $\cL^d$ as well as $\cL$ are nonlinear 
maps.
\end{examp}

Let us pass to non-abelian case, i.e. we replace
$\A_{abel}$ by $\A$, where $\A$ is an arbitrary
nonabelian $C^*$ algebra.
We remind that $\A$ can be considered
as a noncommutative analogue of the space of bounded
continuous functions, i.e. bounded continuous
functions over a ``quantum plane''.
Here, we are using the phrase ``quantum plane'' in the sense 
of Manin (see \cite{Manin}).
Let us repeat the argument given at the beginning of this 
subsection but now applied for the quantum plane.
We can expect that the maps of the form
\begin{equation}
a \mapsto \sum_n c_n a^n
\label{H10}
\end{equation}
where $a \in \A$, $c_n$ is a number,
have arisen as a result of lifting nonlinear
maps defined over the quantum plane
(cf. Theorem 1, Section IV).
In other words, {\it it is our hypothesis that the maps of the
form (\ref{H10}), or more generally completely
positive nonlinear maps on a $C^*$-algebra
can be considered as a result of (nonlinear) lifting nonlinear
maps defined on the quantum plane}. This would mean that the quantization
of a Newtonian system could lead to genuine
nonlinear quantum map.

\section{Quantum Lyapunov exponents.}
In this Section we develop the theory of quantum Lyapunov exponents
(cf. \cite{MK1}, \cite{MK2}).
Let $\A$ be a $C^*$-algebra with the unit $\1$, $\tau : \A \to \A$
a quantum map, i.e. 
\begin{enumerate}
 \item $\tau$ is positive: $\tau(A^*A) \ge 0$
for all $A \in \A$,
 \item  $\tau(\1) = \1$.
\end{enumerate}
The pair $(\A, \tau)$ consisting of a $C^*$-algebra $\A$
and a quantum map $\tau$ will be called the
quantum system.
We propose the following generalization 
of the  classical Lyapunov exponent:

\begin{defn}
\label{qchaos1}
Let $(\A, \tau)$ be a quantum system where
$\tau$ is differentiable (in the Fr\'echet sense) map.
Let  $D_x{\tau}^n(y)$
denote the directional derivatives of ${\tau}$ 
composed with itself $n$ times at a point $x$ in a direction
$y$. 
Then, the limit
\begin{equation}
\label{L1} 
\lambda^q(\tau ; x, y) = \lim\limits_{n\to\infty}\,{1\over
n}\, \log\Vert(D_x{\tau }^n)(y)\Vert \quad (\equiv \lambda^q), 
\end{equation}
whenever exists, will be called the quantum exponent.
\end{defn}
Let us comment on this definition.
Firstly, it is clear that
(\ref{L1}) can be generalized for continuous
dynamical systems in an obvious way (see Section III).
Secondly, one can study the following
weaker version of quantum characteristic exponent:
\begin{equation}
\label{L2} 
\lambda^q_{\mu}(\tau ; x, y) = \lim\limits_{n\to\infty}\,{1\over
n}\, \log| \mu \bigl((D_x{\tau }^n)(y) \bigr) |
 \quad (\equiv \lambda^q_{\mu}), 
\end{equation}
where $\mu$ is a state on $\A$.
However, as we are basically interested in quantum
chaos it is enough for us to look for the largest 
Lyapunov exponent. As $\lambda^q$ plays such the role
(see properties of quantum exponents) we shall
restrict ourself to definition involving the norm.
Thirdly,
the noncommutative generalization (\ref{L1}) of characteristic exponent
is not the only one possibility. Namely,
within the Connes's noncommutative geometry, \cite{Connes}, the required
differential structure can be introduced by derivations 
(see Appendix for definitions) $\{ \delta \}$
of the algebra $\cal A$ associated with a physical system.
Then using basically the same idea (see \cite{ENTS}) one can define
\begin{equation}
\label{L3}
\Lambda^q=\lim_{t \to \infty} {1\over t}\log \Vert \delta
\tau_t(x) \Vert 
\end{equation}
where $x\in {\cal D}(\delta )$ and $\tau_t$ stands for
(continuous) dynamics of the system.
Moreover, we should mention that
the presented two examples of definitions of quantum Lyapunov
exponents do not exhausted all possibilities 
(cf. \cite{SlomZycz}, \cite{V}, \cite{V1}).
However, studying the properties of quantum exponents
as well as looking for quantum counterparts
of chaotic systems (see Section VII) we have found
$\lambda^q$  as the best suited.
Finally, the slight modification of the definition
of Lyapunov exponents described in Remark 1 (see Section III)
can be also done for quantum exponents.
In fact, we shall use such the modification
in our examples (5) and (6) (see Section VII).

Let us turn to the question of existence of the limit
in the definition of $\lambda^q$. In other words,
to present a
convincing argument that $\lambda^q$ is well defined we should give
an example of sufficient conditions implying the existence of
the limit in (\ref{L1}). Let us assume:
\begin{enumerate}
\item $\tau$ is a completely, in general nonlinear, 
positive map (see Appendix for appropriate definitions). In fact, this is
the most important assumption for the next theorem.
Namely, this condition implies a smoothness
of the quantum map $\tau$; even one can say that the complete
positiveness implies a kind of Laurent expansion for $\tau$, 
(see Appendix for details).

\item $\Vert{\tau }^l (0)\Vert \le C_1 $
for all $l\in N$ and some positive $C_1$, and
\begin{equation}
\label{L4}
{\it\Theta}_{\tau } = \{ x \ne 0 : \Vert{\tau }^l (x)\Vert \le
C_2\Vert x\Vert + \Vert{\tau }^l (0)\Vert \} \ne \emptyset 
\end{equation}
for some positive $C_2$ and all $l\in N$.

\item  Finally
\begin{equation}
\label{L5}
\Vert D_x{\tau }^k(y)\Vert > C^k(x,y)
\end{equation}
for some positive $C(x,y)$ and all large $k\in N$.
\end{enumerate}
Under the above assumptions one can prove (cf. \cite{MK2}):
\begin{theo}
\label{qlap}
Let $\tau : {\cal A} \to {\cal A}$ be a map
such that the assumptions (1), (2), (3) are satisfied.
The following limit 
\begin{equation}
\label{L6}
{\lim\limits_{n\to\infty}\,{1\over n}\,\log\,\Vert\,
D_x\,{\tau }^n\,(\,y\,)\, \Vert} 
\end{equation}
 exists for $x\in\,{\it\Theta}_{\tau }.$
\end{theo}

The just given theorem clearly shows that the notion
of quantum characteristic exponent $\lambda^q$ is well defined
for a large class of dynamical systems.
Therefore, having the well-established notion of quantum counterpart 
of Lyapunov exponents $\lambda^q$ let us turn to
the description its basic properties.
We have proved (see \cite{MK2}, \cite{MK3})
\begin{itemize}
\item \qquad ${\lambda }^q(x, y) = {\lambda }^q(x, ay)$\quad
for \quad a $\in {\bf R} \setminus \{ 0 \}. $

\item Note that since the map $y \to  D_x {\tau }^n (y)$ is linear
 one , it is 
natural to set
\begin{equation}
{\lambda }^q(x, 0) = - \infty. 
\end{equation}

\item Let ${\lambda }^q(x, y) > {\lambda }^q(x, z) > - \infty $
and, additionally, let $\tau $ satisfy assumption (iii) of the
theorem in the 
direction $y + z$. 
Then:
\begin{equation}
{\lambda }^q(x, y + az) \le {\lambda }^q(x, y),
\end{equation}
where $a \in {\bf R}.$

\item  The function $y \to {\lambda }^q(x, y)$ as the limit of 
continuous functions (in $y$) is, in general, the Baire function of
type I. In particular, the set
$\{ y \vert y \to {\lambda }^q(x, y)$ is a continuous function$\}$
is dense in ${\cal A}$.

\item Let $\cal A$ be the $C^*$-algebra generated by a fixed
self-adjoint operator and identity on some Hilbert space, i.e.
$\cal A$ is an abelian $C^*$-algebra. Furthermore, let
$\tau : \A \to \A$ be a smooth map while
$\phi$ a state on $\A$.
Note that such 
dynamical system $({\cal A}, \tau, \phi)$ should be 
considered as the classical one. Then, if some mild technical 
conditions are met (cf. \cite{MK3}) Definition \ref{qchaos1} leads to 
the largest classical characteristic exponent. 
\end{itemize}

It is important to observe that the above 
listed properties of $\lambda^q$ 
 are reminiscent to that of classical characteristic
exponents (cf. \cite{AC}, \cite{Pesin}).
Therefore, {\it we conclude that the notion 
${\lambda}^q$ is the well defined
 quantum counterpart of (the largest) characteristic exponent}.

\section{Examples.}
In this Section we want to study concrete
models of dynamical systems. In particular, we are interested in
examples of systems with irregular time evolution, i.e.
with systems with positive quantum exponent.
Moreover, we shall consider examples 
with analytically calculated exponents.
Unless otherwise stated the existence of quantum
Lyapunov exponent for each example was proved separately. 
Let us begin with the simplest model.

{\bf (1)} A dynamical system composed of $N$-level quantum
system with dynamics given
by a~linear dynamical semigroup $S_t$ of contractions. The
assumption of local 
nonexpansiveness is always satisfied but the condition of variability is 
satisfied for some directions. In general, ${\lambda }^q(x,y)
 \in (- \infty , 0)$.
As an illustration let us consider $S_t x = e^{-{\rm \lambda }t}
x$, where $t \ge 0$ and ${\rm \lambda} > 0$. Then ${\lambda
}^q(x, \delta x) = -{\rm \lambda }.$ This means the stability of
the considered semigroup dynamics.

{\bf (2)} A dynamical system with dynamics given by Hartree -
type evolution equation (cf. \cite{AM}):
\begin{equation}
\label{E1}
{d\over dt} {\rho }_t = -i[H({\rho }_t ), {\rho }_t ] 
\end{equation}
\begin{equation}
{\rho }_t {\vert }_{t=0} = {\rho }_o ,
\end{equation}
where $H({\rho }) = [Tr ({\rho }Q)]Q$ with $Q = Q^*$,
and $\varrho$ is a density matrix on a finite dimensional
Hilbert space $\cal H$. Let us
recall that this type of
equation can be obtained in the mean field limit for interacting
quantum systems. The explicit solution of (\ref{E1}) is given by 
\begin{equation}
\label{E2}
{\Phi }_t ({\rho }_o ) = \exp {[-iTr(Q{\rho }_o )Qt]} {\rho
}_o \exp {[iTr(Q{\rho }_o )Qt]}. 
\end{equation}
Thus taking the discrete time in (\ref{E2}) and putting ${\tau} =
{\Phi }_1$ in (\ref{L1}) we get:
\begin{equation}
\label{E3}
\lim\limits_{n\to\infty}\,{1\over n}\,\log \Vert \,-i
[Tr({\delta } {\rho }Q)Qn, {\rho }] + {\delta }{\rho }\,\Vert =
0. 
\end{equation}
In (\ref{E3}) $[\cdot,\cdot]$ denotes the commutator. 
Clearly, assumptions of our theorem are
satisfied. As $\lambda^q =0$, it follows that 
Hartree-type evolution (\ref{E1}) is 
an example of the regular motion.

{\bf (3)} As the next example let us consider the 
dynamical system $(\B,\Phi )$ where $\B$ denotes all linear
operators on a finite dimensional Hilbert space while
the map $\Phi$ is given by
\begin{equation}
\label{E4}
{\Phi }({\rho }) = {\rho }^2 
\end{equation}
for ${\rho }^* = {\rho }$ from unit ball in $B(\cal H)$.
We want to stress that this example is rather
``pure'' mathematical one in the sense that it is well
defined mapping as well as the ``second order'' term
in the expansion of completely positive nonlinear
maps on $C^*$-algebra (cf. Appendix). 
But, it is difficult
to provide a (physical) Hamiltonian leading to the considered map.
On the other hand, let us recall that the ``square'' is 
the basic ingredient of the logistic map - the famous
paradigm of (classical) chaotic systems. 
So, if one assumes that
the nonlinear version of Koopman's construction
can be applied to nonlinear quantization (cf. Section IV
and subsection Vc) one can expect to find the physical
interpretation of this map. Finally we want to
add that the transformation of this type was used by Gisin and his coworkes
(cf. \cite{Gisin}) for the study of the theory of quantum 
computers. So one can conclude that such systems may
be of great significance for a general theory.

Having the clear motivation let us turn
to an examination of its properties.
It is easy to observe 
(cf. Euler theorem) that 
\begin{equation}
\label{E5}
\Vert\,D_{\rho }{\Phi }^n ({\rho })\,\Vert = 
2^n\,{\Vert\,{\rho }\,\Vert }^{{2}^n -1} 
\end{equation}
for ${\rho } \not= {\bf 1}$ and ${\rho } \not\in$
Projectors($\cal H$).
To prove (\ref{E5}), it is enough to note that
$||\rho|| = sup_{\lambda \in spectrum(\rho)} |\lambda|$. 
Clearly, the assumptions of the theorem are satisfied. 
Consequently, (\ref{E5}) leads to:
\begin{equation}
\label{E6}
{\lambda }^q ({\rho }, {\rho }) = \lim\limits_{n\to\infty}\,
{1\over n}\,(({2}^n -1) \log \Vert\,{\rho }\,\Vert + n\,\log 2 ).
\end{equation}
As a conclusion, this mathematical example shows that 
the quantum exponent ${\lambda }^q$
can also be positive (put ${\rho }:\,\Vert \rho \Vert = 1$, 
then ${\lambda }^q ({\rho }, {\rho }) = log2 >0)$.
Thus the quadratic map, in some regions, can exhibits
the irregular form of the ``motion''.

\vspace{10mm}

In the next three examples we reexamine dynamical
systems which were semiclassically treated recently.
Namely, using a kind of approximation (cf. our discussion
following (\ref{H5}), it was shown
that these models of quantum origin display
signatures of chaotic motion. Now we want
to treat these systems in pure quantum way and compare results.
We want to add that in these examples
the Hilbert space $\cal H$ of a system {\it is 
not assumed to be of finite dimension.}

\vspace{5mm}

{\bf (4)} Let us consider a quantum two--level system
interacting with a single mode of the electromagnetic field (cf.
\cite{FE}, \cite{KM1}). The Hamiltonian for a such system can be taken as
\begin{equation}
\label{E7}
H = {1\over 2}\hbar\omega_0\sigma_z + \hbar\omega (a^{\dag} a 
+ {1\over 2}) + \hbar\lambda_0\sigma_x (a + a^{\dag} ) 
\end{equation}
where  $\sigma_x ,\, \sigma_z$ are Pauli matrices and    
$a^{\dag}$ and $a$ stand for bose creation and annihilation operators. 
The energy separation of the two--levels is equal to
$\hbar\omega_0$ while the frequency 
of the radiation mode is $\omega$. The time evolution of the system is given 
by the following operator equation
\begin{equation}
\label{E8}
\dot A = {i\over\hbar} [\,H\, ,\,A\,] 
\end{equation}
for an observable $A$, where the dot denotes a time derivative.
Let us remark that the Heisenberg equation (\ref{E8}) gives for this
model first order coupled nonlinear operator equations (cf. \cite{FE}):
\begin{eqnarray}
\dot \sigma_x &= - \omega \sigma_y& \\
\dot \sigma_y &=\omega_0\sigma_x -2\lambda\sigma_z (a + a^{\dag})&\nonumber \\
\dot \sigma_z &= 2\lambda\sigma_y (a + a^{\dag})& \nonumber \\
\dot{(a + a^{\dag} )} &= -i\omega (a - a^{\dag} )& \nonumber \\
\dot{(a -a^{\dag})} &=-i\omega (a +a^{\dag})-2i\lambda\sigma_x &\nonumber \\ \
\nonumber \end{eqnarray}
Although the Hamiltonian $H$ depends, in general, on the time $t$, 
one can solve (\ref{E8}) and subsequently compute the quantum
exponents ${\lambda}^q$ (see \cite{KM1}). In particular one has:
\begin{equation}
\label{E9}
\lambda^q(\sigma_i ,\sigma_j ) = 0 
\end{equation}
for $i,j = 1, 2$, where $\sigma_1 = \sigma_x$ and $\sigma_2 =
\sigma_z$. Consequently, this quantum optic model
exhibits quantum regular evolution evolution.
Let us add that using some approximations a
signature of chaos was shown (cf. \cite{FE}).

{\bf (5)} This example is the quantum counterpart of a parametrically kicked
nonlinear oscillator (cf. \cite{KM2}). Its hamiltonian $H$ is of the form
\begin{equation}
\label{E10}
H = {\chi \over 2}(a^{\dag})^{2}a^2 + i{\hbar}{\kappa \over2}
[(a^{\dag})^{2} -a^{2}]\cdot\sum_{n=-\infty}^{+\infty} 
\delta (t - nt_0 ) 
\end{equation}
where $\chi$ and $\kappa$ are constants characterizing the system,
$a^{\dag}$ and $a$ stand for bose creation and annihilation operators, and 
$t_0$ is the period of free evolution (i.e. the
evolution described by $H_{NL} = {\chi \over
2}(a^{\dag})^{2}a^2$). 
Let us remark that $H_{NL}$ gives the nonlinear part of the evolution.
Obviously, ${\delta}(\cdot )$ stands now for the Dirac distribution. Then,
the Heisenberg equation of motion,
\begin{equation}
\label{E11}
\dot A ={i\over {\hbar}}[H,A] 
\end{equation}
can be solved in
terms of self--adjoint operators $\Phi$ and $\Pi$, 
where $\Phi$ and $\Pi$ are defined by:
$a = \Phi + i\Pi $. The solution of the
Heisenberg equation for the considered model can be written as:
\begin{equation}
\label{E12}
\left( \matrix{{\Phi}(t_{n}^{+})\cr
                 {\Pi}(t_{n}^{+})\cr}
        \right)
        = e^{-i{\mu}\over 2}
        \left( \matrix{ e^r \cos {\mu}B_0 & e^r \sin {\mu}B_0   \cr
                -e^{-r} \sin {\mu}B_0 & e^{-r} \cos {\mu}B_0\cr}
         \right) 
         \left( \matrix{{\Phi}(t_{n-1}^{+})\cr
                 {\Pi}(t_{n-1}^{+})\cr}
        \right) 
\end{equation}
where ${\mu} = {\chi}t_0$, $r$ the effective constant for kicks,
 and $B_0 = a^{\dag}a -{1\over 2}=
 ({\Phi}^2 + {\Pi}^2 -1)$.

\vspace{5mm}

Let us remark that (\ref{E12}) gives a nonlinear evolution since the
matrix on the right hand side of (\ref{E12}) is the nonlinear function of
operators $\Phi$ and $\Pi$.
We want to study the stability properties of the evolution (\ref{E12}).
We restrict ourself to an examination of the Lyapunov
instabilities of the quadrature components of the electric
field during the time evolution given by the formula (\ref{E12}).
To do this let us define the quadrature operators:
\begin{equation}
\label{E13}
{\Phi}^{\epsilon} = {1\over 2}[e^{i{\epsilon}}a +
e^{-i{\epsilon}} a^{\dag}] 
\end{equation}
and
\begin{equation}
\label{E14}
{\Pi}^{\epsilon} = {1\over 2i}[e^{i{\epsilon}}a -
e^{-i{\epsilon}} a^{\dag}] 
\end{equation}
where $\epsilon \in [0, 2\pi]$.
Let us note that the operators ${\Phi}^{\epsilon}$ and
${\Pi}^{\epsilon}$ are related to the amplitude components of
electric field. At this point in order to avoid any confusion
some explanation is necessary. Namely, the Hamiltonian (\ref{E10})
leads to {\it the nonlinear evolution map} (\ref{E12}). However,
from physical point of view, {\it an examination of the stability
properties of dynamical map does not allow any change of
annihilation (so also creation) operators nor any modification of
the fixed Hamiltonian.} This makes the big difference between
the treatment of classical and quantum models.
Namely, in classical models to study stability properties
of dynamical maps on the phase space (for the special choice
of observables, cf. Sections II and III) we can 
take a slightly different coordinate and momentum.
Here, creation and annihilation operators play the role
of coordinate and momentum.
Again, we made a very particular choice of observables.
The result was nonlinear operator-valued equations.
Therefore,  to 
find {\it physical examples with positive quantum exponents
$\lambda^q$} and to keep the uniqueness of creation and annihilation
operators
we are examining the quadrature operators. In other
words, the quadrature operators are examples of dynamical
variables in our model which can be varied without any change
of $a \quad (a^{\dag})$ and the Hamiltonian.
Thus we shall study quantum exponents in the sense
of Remark 1 (see Section III).

To compute quantum exponents for quadrature operators in the
considered model we should
find an explicit formula for the norm of $\Vert
D_{\epsilon}{\tau}^n (y)\Vert$, where in the example considered,
by $\tau$ we denote the time evolution of the system between two
successive kicks as well as the effect of the first kick. 
The variable 
$y$ stands for an element of the set $\{ \Phi ,\Pi \}$. Clearly,
${\tau}^n$ denotes $\tau$ composed with itself $n$ times.
$D_{\epsilon}$ is the derivative with respect to the phase angle
$\epsilon$ which describes a ``rotation'' in $({\Phi}, {\Pi})$
variables and which is used in the definition of quadrature
operators. But, since $a^{\dag}$ and $a$ are unbounded operators, in
order to compute $\Vert
D_{\epsilon}{\tau}^n (\cdot)\Vert$ we should introduce 
$({\Phi}_{\delta}, {\Pi}_{\delta})$, the
cut--off in the $({\Phi}, {\Pi})$ variables (see \cite{KM2}
for details). Then,
replacing the original dynamics by its well defined approximation
and changing the variables $({\Phi}, {\Pi})$ to $(\tilde{\Phi},
\tilde{\Pi})$ one can find such set of parameters
(${\chi}$,~${\kappa}$,~$t_0$) that ${\lambda}^{q}(\tilde{\Pi}) >
0.$ It is remarkable that this result does not
depend on the performed cut-off.
In other words, for large enough $r$ ($r$ is the effective
constant for kicks) and some values of ${\mu}= t_0 \kappa$, 
the quantum characteristic exponent ${\lambda}^{q}$ for quantum variable
$\tilde{\Pi}_{\delta}$ is strictly positive. To comment on this
result let us note that in polyparametric
cases physics as well as mathematics allow numerous combinations
of stability in certain directions and irregularity in others.
Consequently, in such the cases
one can expect chaotic features of the 
evolution of the model for
some values of ${\chi}$,~${\tau}$, $r$, ~${\kappa}$ and regularity for
others. Here, we got a confirmation of such the behavior 
(${\lambda}^{q}(\tilde{\Phi}_{\delta})$ can be negative
for some values of parameters $\kappa, \chi,$ and $r$
and positive for other values of parameters). Moreover,
let us add that the approach based on some approximations
(cf. remark preceding example (4))
to the considered problem gives
the similar result (see \cite{Mil}, \cite{MH}). 
We conclude that this model
exhibits the hyperbolic type of instabilities.

{\bf (6)} In the last example we consider
the squeezed light in a nonlinear medium having
the second--order susceptibility (${\chi}^{(2)}$). An
analysis of such system leads to the
following equations 
(see \cite{Y}, section 11):
\begin{equation}
\label{E15}
{d\over dz}a(z)=ka^{\dag}(z), 
\end{equation}
\begin{equation}
\label{E16}
{d\over dz}a^{\dag}(z)=\overline{k} a(z),
\end{equation}
where $a(z)$ is the annihilation operator, $k$ is the coupling
constant and $\overline{k}$ stands for 
the complex conjugation of $k$. The dependence 
of $a$ on $z$ is a result of the one dimensional
propagation of the electric field along the
z-axis.
Obviously, the equations (\ref{E15}), (\ref{E16}) lead to
\begin{equation}
\label{E17}
{d^2\over dz^2}a(z) = |k|^2 a(z). 
\end{equation}
It is easy to observe that the same equation can be derived from
the Hamiltonian
\begin{equation}
\label{E18} 
H=\hbar \omega a^{\dag}a+i\hbar{\kappa \over 2}\bigl(
a^{\dag 2}-a^2\bigr)
\end{equation}
where $\omega = 0$. Such a Hamiltonian would appear 
if we take into account an oscillatory time 
dependence of $a$, and then eliminate the free evolution by the
interaction picture.
The solution to equations (\ref{E15}), (\ref{E16}) is
\begin{equation}
\label{E19} 
a(z) = \cosh (|k|z)a(0) + {k\over |k|}\sinh (|k|z)a^{\dag}(0).
\end{equation}
It is clear that we can apply our
definition of Lyapunov exponent $\lambda^q$ to analysis of 
stability of the propagating squeezed light. 
The quadrature operators are taken in the form
\begin{equation}
P_\alpha(z)=
{1\over 2}\Bigl(e^{i\alpha}\,a(z)
+e^{-i\alpha}\,a^{\dag}(z)\Bigr)
\end{equation}
\begin{equation}
Q_\alpha(z)={1\over 2i}\Bigl(e^{i\alpha}\,a(z)
-e^{-i\alpha}\,a^{\dag}(z)\Bigl)
\end{equation}
where $\alpha \in [0, 2\pi]$.
In particular, we have
\begin{equation}
P_\alpha(z)={1\over 2}\Bigl([e^{|k |z}\sin\alpha 
+ e^{-|k |z}\cos\alpha ]a(0) 
\end{equation}
\begin{equation}
+ [-e^{|k |z}\sin\alpha + e^{-|k
|z}\cos\alpha]\,a^{\dag}(t)\Bigr), 
\end{equation}
An easy calculation leads to  
\begin{equation}
\label{E20}
\lambda^q=\lim_{n\rightarrow\infty}
\lim_{z\rightarrow\infty}{1\over z}\ln
\Bigl\Vert{dQ^n_\alpha(z) \over d\alpha}\Bigr\Vert = |k | \geq
0
\end{equation} 
where we put for simplicity ${k\over |k|} = -1$,
$Q^n_{\alpha}$ indicates the cut off in $Q_{\alpha}$ 
(since $a$ and $a^{\dag}$ are unbounded operators
we again have to apply cut-off). 

It is remarkable, see (\ref{E20}), that this result
does not depend on the cut-off.
Hence, this is another quantum optic model with very unstable dynamics.
It is important property of this model that the classical
Lyapunov exponent calculated for the trajectory
\begin{equation}
\label{E22}
t\mapsto \langle w|Q_\alpha(t)|w\rangle 
\end{equation}
has the same value as its quantum counterpart; in (\ref{E22}) $w$ denotes
a standard coherent state $a(0)|w\rangle=w|w\rangle$.

\section{Conclusions.}

We have shown that Quantum Mechanics allows some
form of nonlinear equations for dynamical maps.
Consequently, we have found a kind of coexistence of
regular and irregular quantum motion;
the coexistence resembling very much that of classical
mechanics. Both forms of nonlinearity, in Classical and
Quantum Mechanics, arose as a result of the special
choice of observables ($\{p,q \}$ for classical systems,
$\{ a, a^{\dagger} \}$ for quantum systems).
However we want to repeat that there is very important 
difference in the study of stability properties
of the corresponding dynamical systems.
Namely, studying a stability of classical maps for such
the choice of dynamical variables one can consider an
arbitrary (small) variation of $p$ and $q$.
Basically we can repeat this argument for quantum maps defined
on the set of states (cf. examples 2 and 3, Section VII).
{\it However, we can not do this for $a$ and $a^{\dagger}$!}
The proper way out of this situation is to study
another class of observables, e.g. 
the quadrature operators (cf. examples 5 and 6, Section VII).
Namely, although the quadrature operators
are built directly from $a$ and $a^{\dagger}$, additionally,
they depend on a parameter (phase).
Consequently, their evolution is described by nonlinear maps
while their variation can be implemented by that of the
parameter (phase). Clearly, this procedure
does not effect the uniqueness of $a$, $a^{\dagger}$.
To summarize we have shown that the 
Quantum Mechanics admits the irregularity of the
evolution of some variables. We consider
the positivity of $\lambda^q$ - so irregular motion -
as the basic signature of unstable dynamics. However, we do not 
identify this signature of chaos with the very chaos
of dynamical system
since to get a complete description of chaotic behavior 
of dynamical maps some additional geometrical 
assumptions on the space of dynamical variables 
as well as ergodic assumptions are necessary
(cf. Section III).

The paper also deals with the related question
of derivation of nonlinear quantum dynamical
maps without resorting to quantization of equations
of motion. We have shown the possibility for nonlinear lifting 
of dynamical maps. 
These investigations were intended as an attempt to
motivate our interest in completely positive
nonlinear maps on $C^*$-algebras.
Our result suggests that there is a chance to obtain
such the maps via ``physical considerations''.
However this question is at present far from
been solved as we do not know the relation between the algebraic
structure $\cW_{\om}$ and the phase space $\Gamma$.

We want to end this paper with the remark 
that the presented results are steps toward
a quantum ergodic theory, where a rigorous
definition of quantum chaos may be established.

\section{Appendix.}
The theory of $C^*$-algebras can be find in the books of
\cite{BrRo}, \cite{KadRin}, \cite{Sakai} or \cite{Tak}.
Here we recall only few basic results which have been used
in the paper.
\begin{itemize}
\item A Banach $^*$-algebra $\A$ is called a $C^*$-algebra
if it satisfies $||a^*a|| = ||a||^2$ for $a \in \A$.
\item Let $\A$ be an abelian $C^*$-algebra, i.e.
$ab=ba$ for any $a,b \in \A$.
A character $\vp$, of $\A$, is a nonzero linear map, $\A \ni a
\mapsto \vp(a) \in \Cn$ such that
\begin{equation}
\vp(ab) = \vp(a) \vp(b)
\end{equation}
foe all $a,b \in \A$. The spectrum $\cP(\A) \equiv \cP$,
of $\A$ is to be the set of all characters on $\A$.
\item If $\vp$ is a character of an abelian $C^*$-algebra
$\A$ then $\vp(a) \in \sigma(a)$, the spectrum
of (the operator) $a$  for all $a \in \A$.
Moreover, $|\vp(a)| \le ||a||$ and $\vp(a^*a) \ge 0$.
\item ({\em Gelfand-Naimark theorem})
Let $\A$ be an abelian $C^*$-algebra and $\cP$
the set of characters of $\A$ equipped with the weak$^*$ topology 
inherited from the dual $\A^*$, of $\A$. It follows that $\cP$
is a locally compact Hausdorff space which is compact
if and only if $\A$ contains the identity.
Moreover, $\A$ is isomorphic to the algebra
$C_0(\cP)$ of continuous functions over $\cP$ which
vanish at infinity.
\item A (symmetric) derivation $\delta$ 
is a linear map from a $*$-subalgebra
${\cal D}(\delta)$, the domain of $\delta$, into $\cal A$
with the properties that
\begin{enumerate}
\item $\delta(A)^*=\delta(A^*), \qquad A\in {\cal D}(\delta)$
\item $\delta(AB)=\delta(A)B + A\delta(B), \qquad A,B \in
{\cal D}(\delta)$
\end{enumerate}
\item {\em Theorem. (\cite{BrRo})} Let $\A$ be a $C^*$-algebra
with identity $\1$, and let $\delta$ be a norm-densely
defined norm-closed operator on $\A$ with domain
${\cal D}(\delta)$. It follows that $\delta$ is the generator
of strongly continuous one-parameter group
of $^*$-automorphism of $\A$ if and only if
${\cal D}(\delta)$ is a $^*$-algebra and $\delta$
is a symmetric derivation.

\vspace{3mm}
 
The listed properties of $\delta$
imply $\1 \in {\cal D}(\delta)$
and $\delta(\1) =0$.

\item (\cite{Bratteli}) Let again $\A = C(\Omega)$
be an abelian $C^*$-algebra.
By the dimension of $\Omega$ we shall mean the Hausdorff
dimension of $\Omega$.
The classification results for derivations of $\A$ can be given
in the following way:
\begin{enumerate}
\item In the case that $dim\Omega =0$
it turns out that all closed derivations
are trivial (i.e. equal to 0) if $\Omega$ is total disconnected.
\item The case $dim \Omega =1$,
$\A = C([0,1])$ has the complete classification.
\item For dimension more than 2 only sporadic results are known.
\end{enumerate}
These results clearly show that in our discussion
of the Hamilton picture (cf. Section II) implicitly we have used 
some additional topological properties of the phase space
$\Gamma$.
\item Let $\A_i$, $i=1,...,k$ and $\cB$ be $C^*$-algebras
and $\A_1 \times \cdot \cdot \cdot  \A_k$
the Cartesian product
of $\A_1,...,\A_k.$
A map $\Phi : \A_1 \times \cdot \cdot \cdot  \times \A_k \to \cB$
is said to be completely positive if, for every $n>0$,
the $n$-square $\cB$-valued matrix 
$[\Phi(a_{1;p,r},...,a_{k;p,r})]_{p,r=1}^k$
is positive whenever the $n$-square $\A_i$-valued
matrices $[a_{i; p,r}]_{p,r}$ are positive for $i=1,...,k$.
\item {\em Theorem (\cite{ACh})} Let $\Phi: \A \to \cB$
be a completely positive  (in general, nonlinear) map
between two $C^*$-algebras $\A$ and $\cB$.
Then, there exist (uniquely) completely positive
maps $\Phi_{m,n} : \A \to \cB$ (m,n = 0,1,2,...)
such that
\begin{enumerate}
\item $\Phi(a) = \sum_{m=0}^{\infty} \sum_{n=0}^{\infty}
\Phi_{m,n} (a)$
\vspace{3mm}
where $a \in \A$ and the series in norm convergence.
\item $\Phi_{m,n}(za) = z^m {\overline z}^n \Phi_{m,n}(a)$
\vspace{3mm}
where $z \in \Cn$ and $a \in \A$.
The map $\Phi_{m,n}(\cdot)$ with the above property is
called $(m,n)$-mixed homogeneous.
\end{enumerate}
\item {\em Theorem (\cite{ACh})}
Let $\Phi : \A \to \cB$ be $(m,n)$-mixed homogeneous
map with $m+n>0$.
Then, there is a completely positive map 
$\phi :   \A_1 \times \cdot \cdot \cdot \times \A_{m+n} \to \cB$
with $\A_k \equiv \A$ $k=1,...,m+n$ such that
\begin{equation}
\Phi(a) = \phi(a,\cdot, \cdot, \cdot,a)
\end{equation}
where $a \in \A$ and $\phi(a_1,...,a_{m+n})$
is multilinear in $(a_1,...,a_m)$
and multi-conjugate-linear in $(a_{m+1},...,a_{m+n}).$
\end{itemize}

\vskip 1 true cm

{\bf Acknowledgments}: 
This paper is written on the basis of lectures given
by the author in the Nagoya University in October 1997.
The author would like to thank 
Luigi Accardi and Nobuaki Obata
for hospitality at Graduate School of Polymathematics of Nagoya University
as well as for suggesting these lectures.
I am grateful to Andrzej Posiewnik for critical
reading the manuscript.
The author would like also to acknowledge
the partial supports of KBN grant
129/PO3/9509 and the Nagoya University grant.


\end{document}